\documentclass[%
 aip,
 jmp,%
 amsmath,amssymb,
 reprint,%
]{revtex4-2}
\usepackage{float,caption,subcaption}
\usepackage{makecell}
\usepackage{graphicx}
\usepackage{dcolumn}
\usepackage{bm}

\begin{document}

\preprint{AIP/123-QED}

\title{Rationally Correcting Impurity Levels Positions Based on Electrostatic Potential Strategy for Photocatalytic Overall Water Splitting}

\author{Dazhong Sun}
\altaffiliation[Also at ]{College of Science, Nanjing University of Posts and Telecommunications, Nanjing, 210023, China}

\author{Wentao Li}
\altaffiliation{Weifang University of Science and Technology, Shouguang, 262700, China}

\author{Anqi Shi}
\altaffiliation[Also at ]{College of Science, Nanjing University of Posts and Telecommunications, Nanjing, 210023, China}

\author{Wenxia Zhang}
\altaffiliation{School of Optoelectronic Engineering, Chongqing University of Posts and Telecommunications, Chongqing 400065, China}

\author{Huabing Shu}
\altaffiliation{School of Science, Jiangsu University of Science and Technology, Zhenjiang 212001, China}

\author{Fengfeng Chi}
\altaffiliation[Also at ]{College of Science, Nanjing University of Posts and Telecommunications, Nanjing, 210023, China}

\author{Bing Wang}
\email[Corresponding author: ]{wb@henu.edu.cn}
\altaffiliation{Institute for Computational Materials Science, Joint Center for Theoretical Physics (JCTP),
School of Physics and Electronics, Henan University, Kaifeng, 475004, China}

\author{Xiuyun Zhang}
\email[Corresponding author: ]{xyzhang@yzu.edu.cn}
\altaffiliation{College of Physics Science and Technology, Yangzhou University, Yangzhou 225002, China}

\author{Xianghong Niu}
\email[Corresponding author: ]{xhniu@njupt.edu.cn}
\affiliation{College of Science, Nanjing University of Posts and Telecommunications, Nanjing, 210023, China}

\date{\today}

\begin{abstract}
Doping to induce suitable impurity levels is an effective strategy to achieve highly efficient photocatalytic overall water splitting (POWS). However, to predict the position of impurity levels, it is not enough to only depend on the projected density of states of the substituted atom in the traditional method. Herein, taking in phosphorus-doped g-C$_3$N$_5$ as a sample, we find that the impurity atom can change electrostatic potential gradient and polarity, then significantly affect the spatial electron density around the substituted atom, which further adjusts the impurity level position. Based on the redox potential requirement of POWS, we not only obtain suitable impurity levels, but also expand the visible light absorption range. Simultaneously, the strengthened polarity induced by doping further improve the redox ability of photogenerated carriers. Moreover, the enhanced surface dipoles obviously promote the adsorption and subsequent splitting of water molecules. Our study provides a more comprehensive view to realize accurate regulation of impurity levels in doping engineering and gives reasonable strategies for designing an excellent catalyst of POWS.
\end{abstract}

\maketitle

\section{Introduction}

Doping engineering is an effective strategy to achieve highly efficient photocatalytic overall water splitting (POWS) because it can not only give rise to new transitions to expand the solar absorption range but also regulate the redox ability of photogenerated carriers in catalysts\cite{pows1,pows2,pows3}. To maximize the assistance of the doping strategy in catalysis, the regulation for positions of the impurity energy level in the band gap should be as accurate as possible. That is, it is crucial to investigate in-depth formation mechanisms of impurity levels. Based on the effective mass theory, Kohn and Luttinger have employed the envelope function approach to analyze the wave function of electrons in impurity states, expanding it with the wave function of substituted atoms in pure structures\cite{klemt}. Kohn-Luttinger's effective mass theory (KL-EMT) has successfully predicted levels of impurity states in h-BN and some two-dimensional carbon nitrides\cite{klemt1,klemt2}. However, out-plane distortion or polarity of some doped structures may lead to a significant change of electrostatic potential around impurity atoms. In these cases, analyzing impurity states only based on the wave function of the pure case is insufficient\cite{klemt}. It does not imply that KL-EMT is incorrect but does lead to a dilemma in applying KL-EMT in two-dimensional or polar materials.

In the last decade, Yang et al. pointed out employing the polarity to enhance the redox potential of photogenerated carriers\cite{yang}. The intrinsic electronic field brought by polarity can realign band edges on different surfaces and strengthen the redox ability of photogenerated carriers\cite{jihua1,jihua2,jihua3}. Moreover, this intrinsic electronic field can spatially separate photogenerated electrons and holes\cite{jihua1}. This effectively alleviates the dilemma in doping engineering: although the impurity level in the band gap can expand the solar absorption range, the redox ability of photogenerated carriers will be reduced. To date, employing polarity has become a common strategy to realize a balance between the high redox capacity and a wide solar absorption range\cite{jihua3,jihua4,jihua5}. The g-C$_3$N$_5$ is a novel two-dimensional carbon nitride regarded as a popular candidate in photo-catalysis because of its ferroelectricity, broad $\pi - \pi$ system, great specific surface area, and ``earth-abundant'' nature of raw material\cite{c3n5,cn,c3n52}. In addition, the low symmetry in planar and vertical directions can bring variable band structures and potential distribution in doping engineering. It has been proved that phosphorus doping can also improve the performance of photo-catalysis in g-C$_3$N$_5$ to achieve the degradation of Rhodamine B\cite{pc3n5}. However, for this kind of out-plane distorted or polarized 2D materials, it is still difficult to accurately regulate impurity levels based on the traditional KL-EMT.

In this paper, we systemically investigate 19 samples of phosphorus doping g-C$_3$N$_5$ (P-C$_3$N$_5$) by first principle calculations. We find that the projected density of states (pDOS) of substituted atoms in the pure g-C$_3$N$_5$ is only one factor affecting the positions of impurity levels. After doping phosphorus, the potential change on the surface and the polarity induce the electrostatic potential gradient, which further brings the change of impurity electrons distribution around the doping location. This is another factor that has to be considered for accurately predicting impurity energy level positions. According to the requirement of redox potential for oxygen evolution reaction (OER) and hydrogen evolution reaction (HER), four kinds of phosphorus doping possess suitable impurity energy levels. Simultaneously, the electrostatic potential difference between the two surfaces of P-C3N5 can be raised up to 0.82 eV, significantly higher than that of pure g-C$_3$N$_5$ (0.30 eV), which strengthens the redox ability of photogenerated carriers. The enhanced surface dipole (from 0.03 e$\cdot$pm to 0.14 e$\cdot$pm after P doping) can also effectively promote the adsorption and splitting water molecules. Moreover, impurity levels in the band gap widen the photo-absorption range from the ultraviolet to visible light. 

\section{Computation details}
\subsection{Density functional theory(DFT) calculation}
The density functional theory (DFT) calculation is by employing the Vienna Ab initio Simulation Package (VASP) with the projector augmented wave (PAW) method\cite{dft,vasp1,vasp2,vasp3,paw}. The cutoff energy of the plane wave was set to 450 eV. The energy and chemical potential of the pristine g-C$_3$N$_5$, adulterants and doped systems were prepared with the Perdew-Burke-Ernzerhof type functional with the generalized gradient approximation (GGA-PBE)\cite{pbe}. For the more accurate analyses of the band edge and the property of solar utilization, we applied the Heyd-Scuseria-Ernzerhof (HSE06) functional on the calculations of band structures and photo-absorption\cite{hse06}. To avoid the interaction between periodic unit cells and the errors of atomic force, total energy, and electrostatic potential caused by boundary conditions, we employed the dipole corrections with the vacuum space set to 20 \AA. Moreover, we used the DFT-D3 dispersion correction method to correct the van der Waals interactions\cite{vdw3}.

\subsection{Formation energy calculation}
The doping models are built by substituting the atom or the insertion with the impurity atom in the pure system of a 2$\times$2 supercell named the replacement type or the interstitial type. The formation energy $E_{form}$ is performed by the following equations\cite{form1,form2}:
\begin{equation}
E_{form} = E_{(P-C_3N_5)} + \mu(X) - E_{(g-C_3N_5)} - \mu(P) \label{Ef} 
\end{equation}
\begin{equation}
\mu(P) = \dfrac{E_{(P_4)}}{4} \label{Ep} 
\end{equation}
\begin{equation}
E_{(g-C_3N_5)} = 5\mu(C)+ 8\mu(N) \label{Ex} 
\end{equation}
where $E_{(P-C_3N_5)}$, $E_{(g-C_3N_5)}$, and $E_{(P_4)}$ present total energies of the P-doped g-C$_3$N$_5$, the pristine g-C$_3$N$_5$, and the P$_4$ molecular in the gas phase, respectively; $\mu(X)$ presents the chemical potential of the substituted atom. In a doped system of interstitial type, $\mu(X)$ is 0. The chemical potential of a single carbon atom $\mu(C)$ or a single nitrogen atom $\mu(N)$ depends on the different forming situations, calculated by equation (\ref{Ex}) and the following equations:
\begin{equation}
\mu(N) = \dfrac{E_{(N_2)}}{2} \label{En} 
\end{equation}
\begin{equation}
\mu(C) = \dfrac{E_{(graphene)}}{2} \label{Ec}
\end{equation}
where $E_{(N_2)}$ and $E_{(graphene)}$ present total energies of the N$_2$ molecular and unit cell of graphene, respectively. In this study, $\mu(C)$ and $\mu(N)$ are calculated by equations (\ref{Ex}) and (\ref{En}) in the N-rich conditions and by equations (\ref{Ex}) and (\ref{Ec}) in the C-rich conditions.

\subsection{Photo-absorption calculation}
Due to the low symmetry in the planar of the g-C$_3$N$_5$, we calculated the coefficient of the photo-absorption in different directions of $x$ and $y$. The coefficient of light absorption $I(\omega)$ is performed by the following equation\cite{optical1}:
\begin{align}
I(\omega)=\dfrac{\sqrt{2}\omega}{c}\left[ \sqrt{\varepsilon_r^2 + \varepsilon_i^2} - \varepsilon_r \right]
\end{align}
where $c$ and $\omega$ are the speed and frequency of the photons, $\varepsilon_r$ and $\varepsilon_i$ present the real and imaginary parts of the dielectric function $\varepsilon(\omega)$, respectively. The matrix element $\varepsilon_i^{\alpha \beta}(\omega)$ of $\varepsilon(\omega)$ is calculated by the following equation\cite{optical2}:
\begin{align}
\varepsilon_i^{\alpha \beta}=&\dfrac{4\pi^2 e^2}{\Omega}\lim_{q\to 0}\dfrac{1}{q^2}\sum_{C,V,\boldsymbol{k}}2\varpi_{\boldsymbol{k}}\delta(\epsilon_{C\boldsymbol{k}}-\epsilon_{V\boldsymbol{k}}-\omega) \times \left< \left. u_{C\boldsymbol{k}+\boldsymbol{e}_{\alpha}q} \right|u_{V\boldsymbol{k}} \right>\left< \left. u_{C\boldsymbol{k}+\boldsymbol{e}_{\beta}q} \right|u_{V\boldsymbol{k}} \right>
\end{align}
where $\Omega$ and $q$ are the volume of the pristine cell and the electron momentum operator, $C$ and $V$ present the indices of conduction and valence bands, respectively. $\boldsymbol{k}$ and $\varpi_{\boldsymbol{k}}$ are the Bloch wave vector and the weight of the corresponding $\boldsymbol{k}$ point, $\boldsymbol{e}_{\alpha}$ is the unit vector in Cartesian coordinates. $u_{C\boldsymbol{k}}$, $u_{V\boldsymbol{k}}$, and $\epsilon_{C\boldsymbol{k}}$, $\epsilon_{V\boldsymbol{k}}$ are the wave-functions and corresponding eigenvalues at the $\boldsymbol{k}$ point.

\subsection{Surface Dipole calculation}
The surface dipole is calculated by integrating the electron density in the corresponding direction. The overall surface dipole distribution in the $z$ direction is calculated by the following questions\cite{dipol1,dipol2}:
\begin{align}
P_0=\int_{z_1}^{z_2} \rho(z)(z-z_0)\mathrm{d}z
\end{align}
where $\rho(z)$ is the smoothed linear density of electrons in the $z$ direction, $z_1$ and $z_2$ are two side of g-C$_3$N$_5$ systems where the electronic density goes to zero, $z_0$ is the effective charge center of the system which calculated by
\begin{align}
z_0=\dfrac{\int_{z_1}^{z_2} z\rho(z)\mathrm{d}z}{\int_{z_1}^{z_2} \rho(z)\mathrm{d}z}
\end{align}
The electrons distribution is calculated by VASP and is performed by VASPKIT code\cite{vaspkit}.

\section{Results and discussions}

\subsection{Properties of pristine g-C$_3$N$_5$}
\begin{figure}[H]
\begin{center}
	\includegraphics[width=13cm]{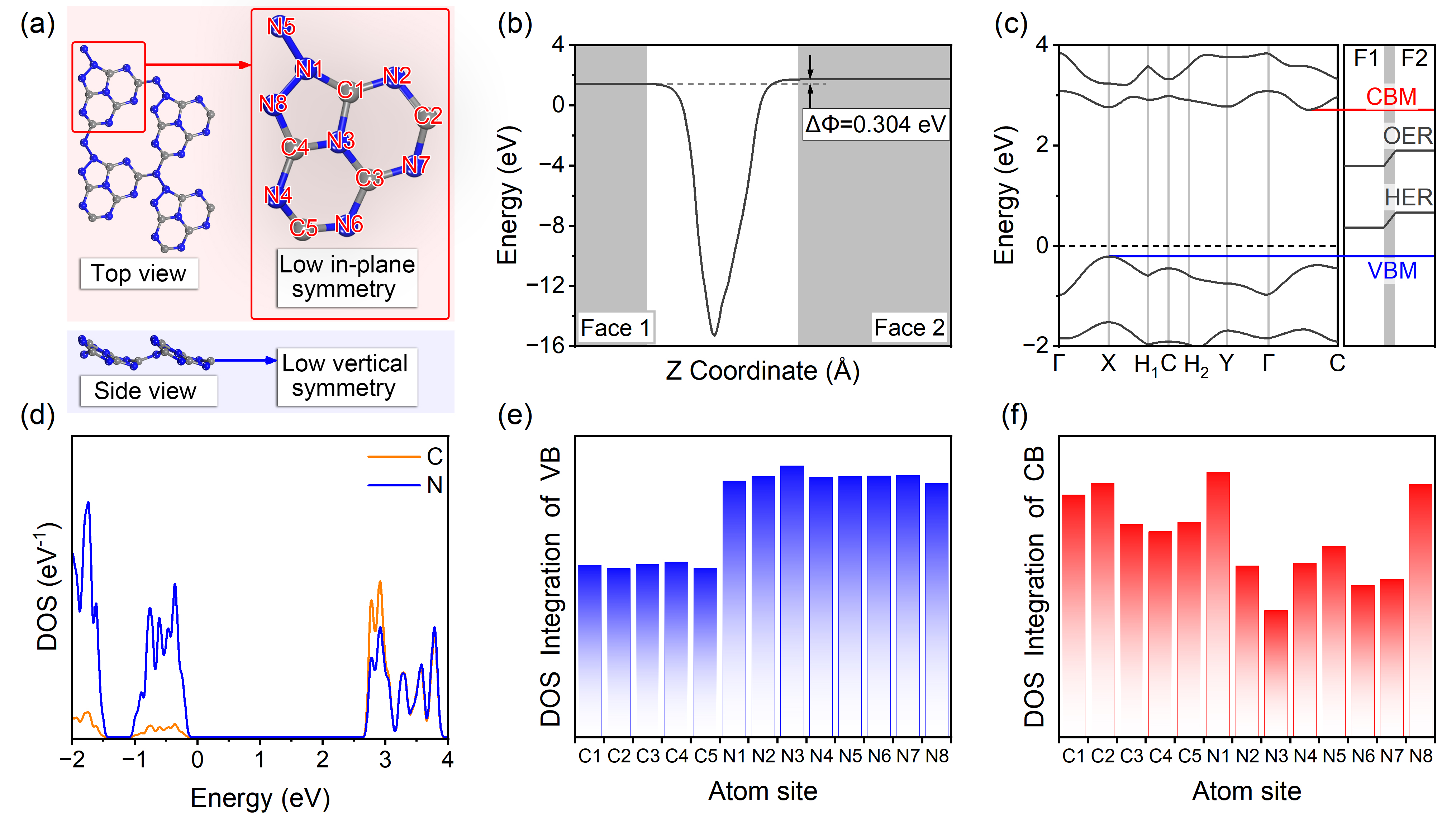}
	\caption {(a) The top and side views of the optimized pristine g-C$_3$N$_5$. We labeled all carbon and nitrogen atoms with the numbers. The gray and blue balls represent carbon and nitrogen atoms, respectively. (b) The variation of electrostatic potential for pristine g-$_3$N$_5$ in  direction. We use the gray block to indicate the faces 1 and 2 of structures. (c) The band structure of the pristine g-C$_3$N$_5$. We provided the corrected potential requirement of OER and HER on the right side of the corresponding structures. Red and blue lines are used to denote CBM and VBM. F1 and F2 present faces 1 and 2 of the doping structures. (d) The projected density of states (pDOS) for pristine g-C$_3$N$_5$. (e) and (f) The integrations of the density of states (DOS) in the regions of valence band and conduction band.} \label{fig1}
\end{center}
\end{figure}

A unit cell of g-C$_3$N$_5$ consists of five carbon atoms and eight nitrogen atoms, including a triazole ring and two triazine rings (seeing Fig.\ref{fig1}(a)). For the sake of the discussion, we numbered each atom of the unit cell and labeled faces 1 and 2 of g-C$_3$N$_5$, and the corresponding labeling was applied to the later discussion.

As shown in Figure 1, the structure of g-C$_3$N$_5$ has a low symmetry in its planar and vertical directions, which induces a weak discrepancy of 0.30 V for the electrostatic potentials in the different surfaces. The wide band gap of 2.95 eV and the suitable band edge energy level (seeing Fig.\ref{fig1}(c)) indicate that the redox ability of photo-generated carriers in g-C$_3$N$_5$ is sufficient to drive the POWS process. However, the large band gap will bring a narrow range of optical absorption spectrum out, which restricts the application of the catalyst. To improve the performance of g-C$_3$N$_5$ in POWS, we explored the doping strategy by inducing phosphorus atom.

\subsection{The position of impurity levels}

\begin{figure}[H]
\begin{center}
	\includegraphics[width=7cm]{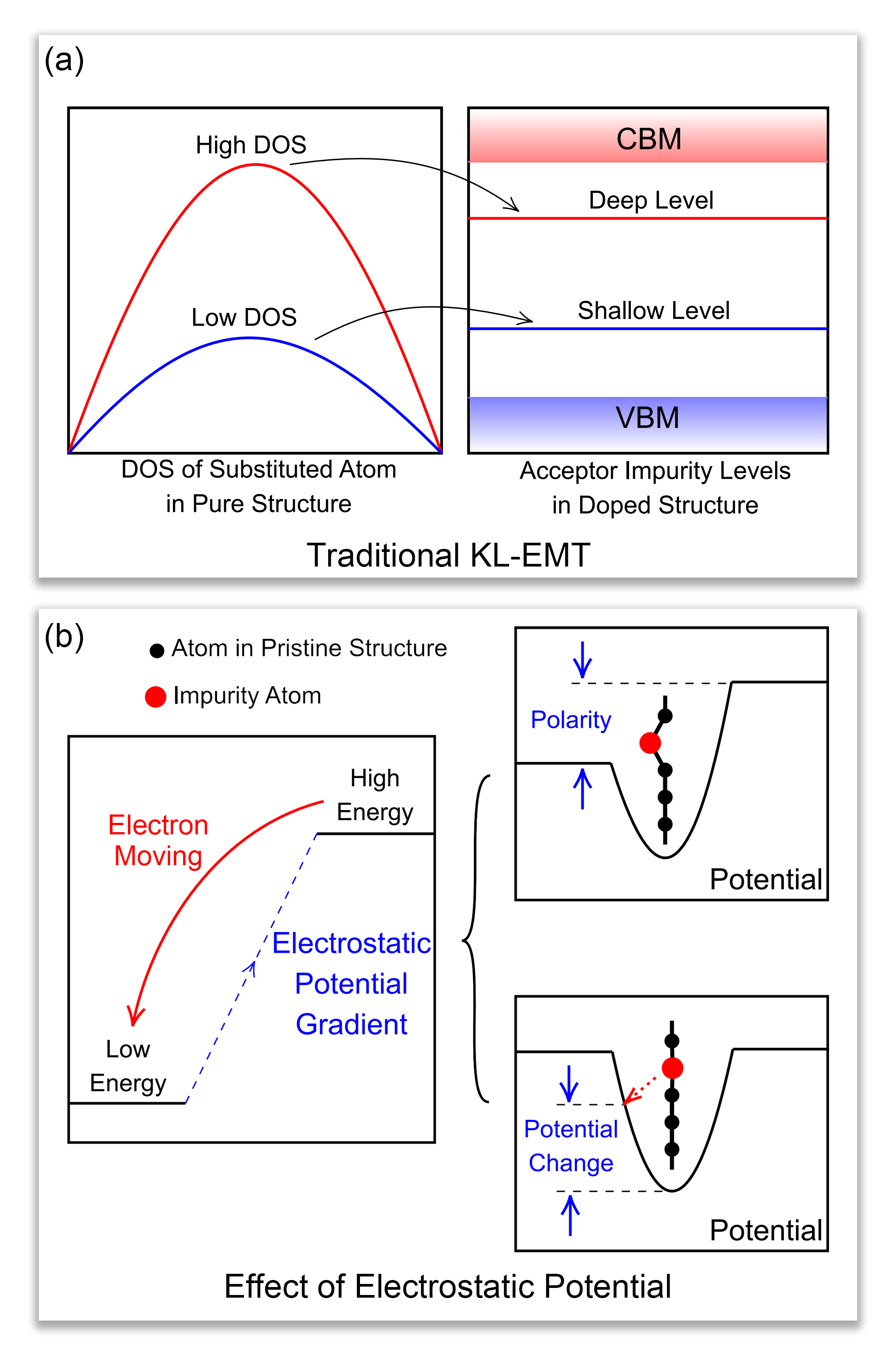}
	\caption {The mechanism of factors affecting the depth of impurity levels. (a) is for traditional KL-EMT. (b) is for the mechanism of electrostatic potential gradient affecting the depth of impurity levels.} \label{fig2}
\end{center}
\end{figure}

Here we use the ``depth" to describe the position of impurity levels: the closer the acceptor(donor) impurity level in the band gap is to the region of conduction(valence) bands, the deeper it is. The depth of impurity levels mainly depends on the spatial density of electrons in corresponding states. In the traditional KL-EMT, the DOS of the replaced atom in a pure system is the main influencing factor of the spatial electron density in impurity states near the doping site in a doped system, and the relationship between them can be described by the following equation\cite{klemt,klemt1,klemt2}:

\begin{align}
\Psi=\sum_{q}\sum_{n}F_{qn}\left(\boldsymbol{r}\right)\varphi_{qn}\left(\boldsymbol{r}\right) \label{kl}
\end{align}

\noindent where the $\Psi$ is the wave function of impurity states, $F_{qn}\left(\boldsymbol{r}\right)$ is the effective mass envelop function solved by the Shindo-Nara effective mass equation, and $\varphi_{qn}\left(\boldsymbol{r}\right)$ is the Bloch wave-function of replaced atom in the pure system\cite{shindonara}. $q$ and $n$ denote the valleys and bands in pristine structure, respectively. Equation (\ref{kl}) indicates that the wave function of the acceptor (donor) impurity states can be expressed as the linear combination of the Bloch wave function from the valleys of the valence (conduction) bands of the pure system. This relationship corresponds to the impurity levels in that the more DOS occupation of a replaced atom in the valence (conduction) band of undoped structure, the deeper the impurity level of the acceptor (donor) dopants (seeing Fig.\ref{fig2}(a)). In this study, we described the DOS occupation of each atom by integrating the DOS in corresponding valence or conduction bands with the following equation:

\begin{align}
\widetilde{N}=\int \rho(E)\mathrm{d}E \label{dos}
\end{align}

\noindent where $\widetilde{N}$ is the DOS occupation of the replaced atom, $\rho(E)$ is the DOS of replaced atom, respectively. The integral covers the region of corresponding valence or conduction bands. As shown in Fig.\ref{fig1}(e) and Fig.\ref{fig1}(f), the low symmetry of g-C$_3$N$_5$ leads to inconsistent DOS contribution for all atoms. But in the region of valence bands, the atom's DOS of the same element is similar. According to KL-EMT, impurity levels induced by acceptor dopants will have similar positions in the band gap.

Traditional theory has roughly predicted the position of impurity levels in the band gap. However, KL-EMT is no longer valid if there is a place where the potential gradient is considerable. The potential gradient consists of two parts: (\romannumeral1) When dopants are on the surface of the structure, there is an electrostatic potential change with the impurity electron around the dopant site. (\romannumeral2) There is an intrinsic electric field with impurity electrons when the doped structure is polar. Both these conditions will bring the electrostatic potential gradient with impurity electrons, which will induce electron movement. At this point, the spatial density of impurity electrons around the dopant site changes, thus affecting the depth of the impurity level in the band gap (seeing Fig.\ref{fig2}(b)).

\subsection{Formation energies and doping types}

\begin{figure}[H]
\begin{center}
	\includegraphics[width=7cm]{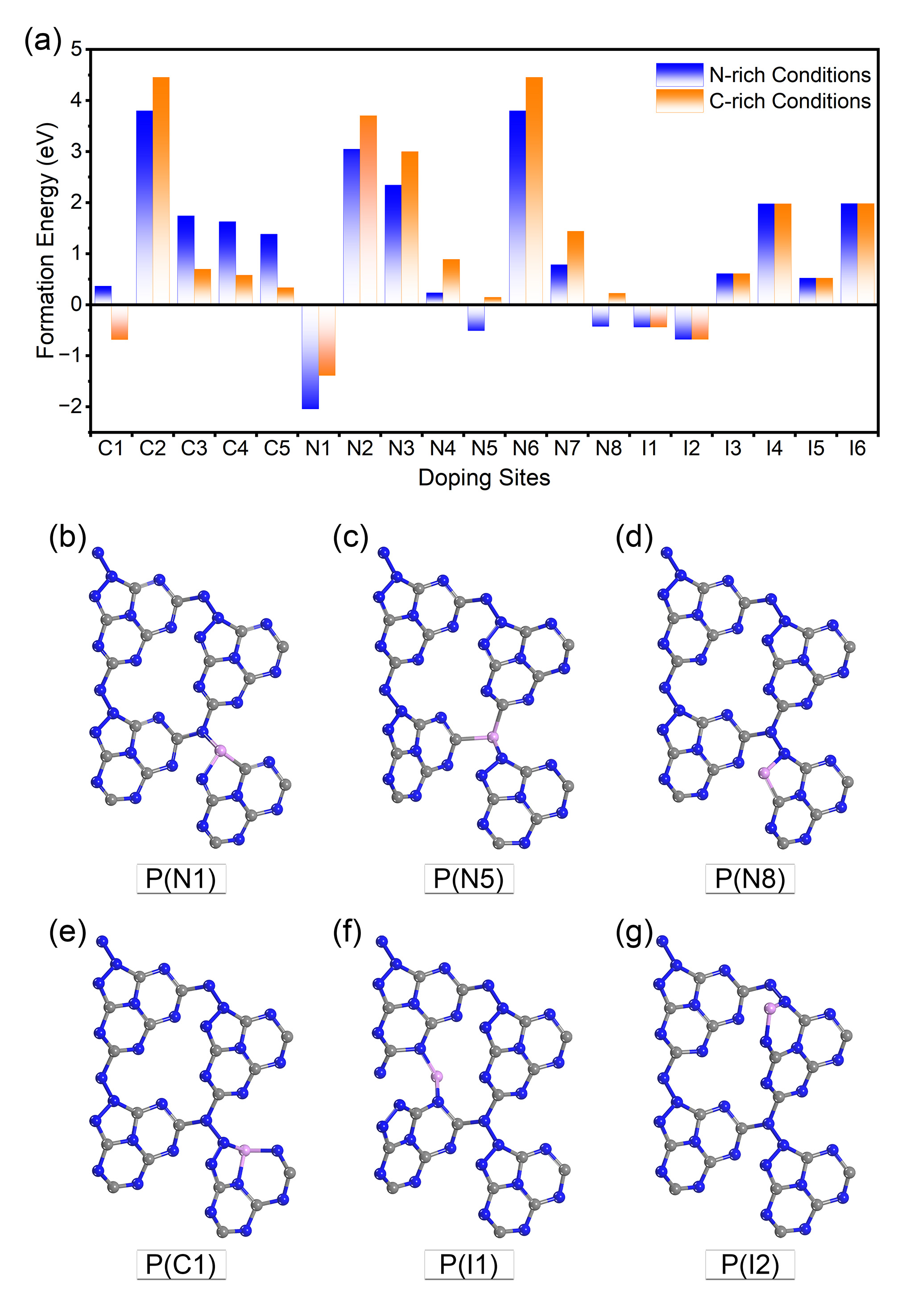}
	\caption {(a) The formation energies of all doping structures. We used blue and orange colors to indicate the different formation energies in C-rich and N-rich conditions, respectively. (b)-(g) The geometry-optimized structures of selected doping structures. The gray, blue, and light purple balls represent carbon, nitrogen, and phosphorus atoms, respectively.} \label{fig3}
\end{center}
\end{figure}

Here we evaluate the thermodynamic stability of doped structures with the formation energy $E_{form}$ and display corresponding results in Fig.\ref{fig3}. For convenience, we label the doped structure as $P(X)$, which $X$ represents the replaced atom corresponding the Fig.\ref{fig1}(a). $P(In)$ represents a doped structure with inserting the impurity atoms into the gap of g-C$_3$N$_5$, which $n$ representing different doping sites. To approach the results in the realistic experimental environment, we calculated the formation energy in two different synthetic environments of C-rich and N-rich conditions. In supporting information, we displayed all doping structures as shown in Figure S1-S3.

As shown in Fig.\ref{fig3}, most of the doped structures have lower formation energies in N-rich environments than in C-rich, but $P(C3)$, $P(C4)$, and $P(C5)$ have lower formation energies in C-rich environments. To ensure that the catalyst for POWS is thermally stable, we set $E_{form}<0~$eV as the standard. There are six doping structures of $P(N1)$, $P(N5)$, $P(N8)$, $P(C1)$, $P(I1)$, and $P(I2)$ are thermally stable. To determine the doping types for the subsequent impurity energy level analysis, we identify the phosphorus atom as a donor or acceptor impurity dopant when it substitutes a carbon or nitrogen atom. In the next subsection, we will analyze the electronic properties of all six doping structures.

\subsection{Electronic properties of selected structures}

\begin{figure}[H]
\begin{center}
	\includegraphics[width=13cm]{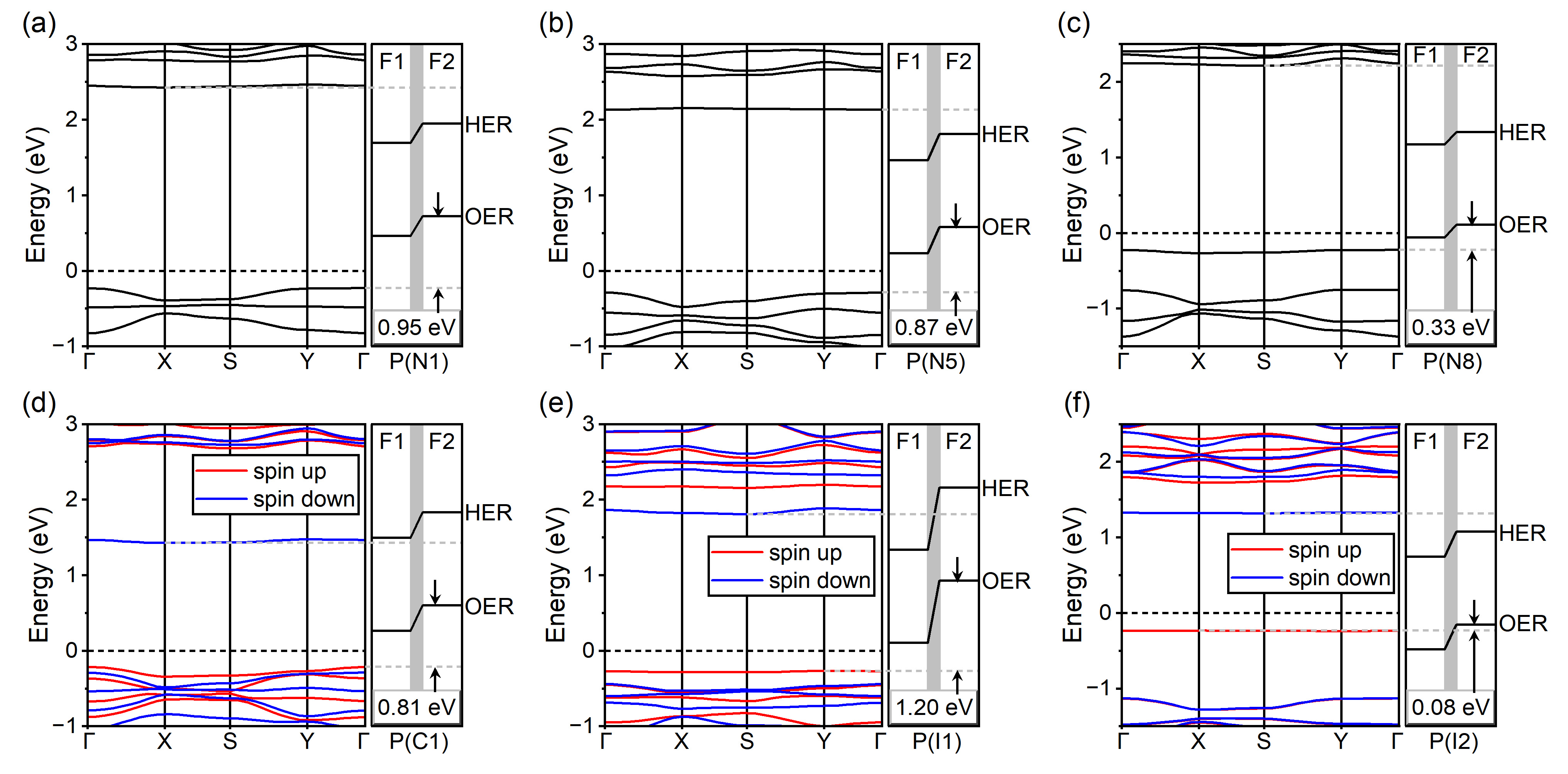}
	\caption {(a)-(f) The accurate band structures of $P(N1)$, $P(N5)$, $P(N8)$, $P(C1)$, $P(I1)$, and $P(I2)$, respectively. We provide the corrected potential requirement of OER and HER on the right side of the corresponding structures. For spin-asymmetric systems such as $P(C1)$, $P(I1)$, and $P(I2)$, red and blue lines present the part of bands with spin up and spin down, respectively. We use short dash lines with light gray to denote the band edge of all doping structures, and the black arrows to point out the gap between VBM and the requirement potential of OER in doping systems. F1 and F2 present faces 1 and 2 of the doping structures.} \label{fig4}
\end{center}
\end{figure}

\begin{figure}[H]
\begin{center}
	\includegraphics[width=13cm]{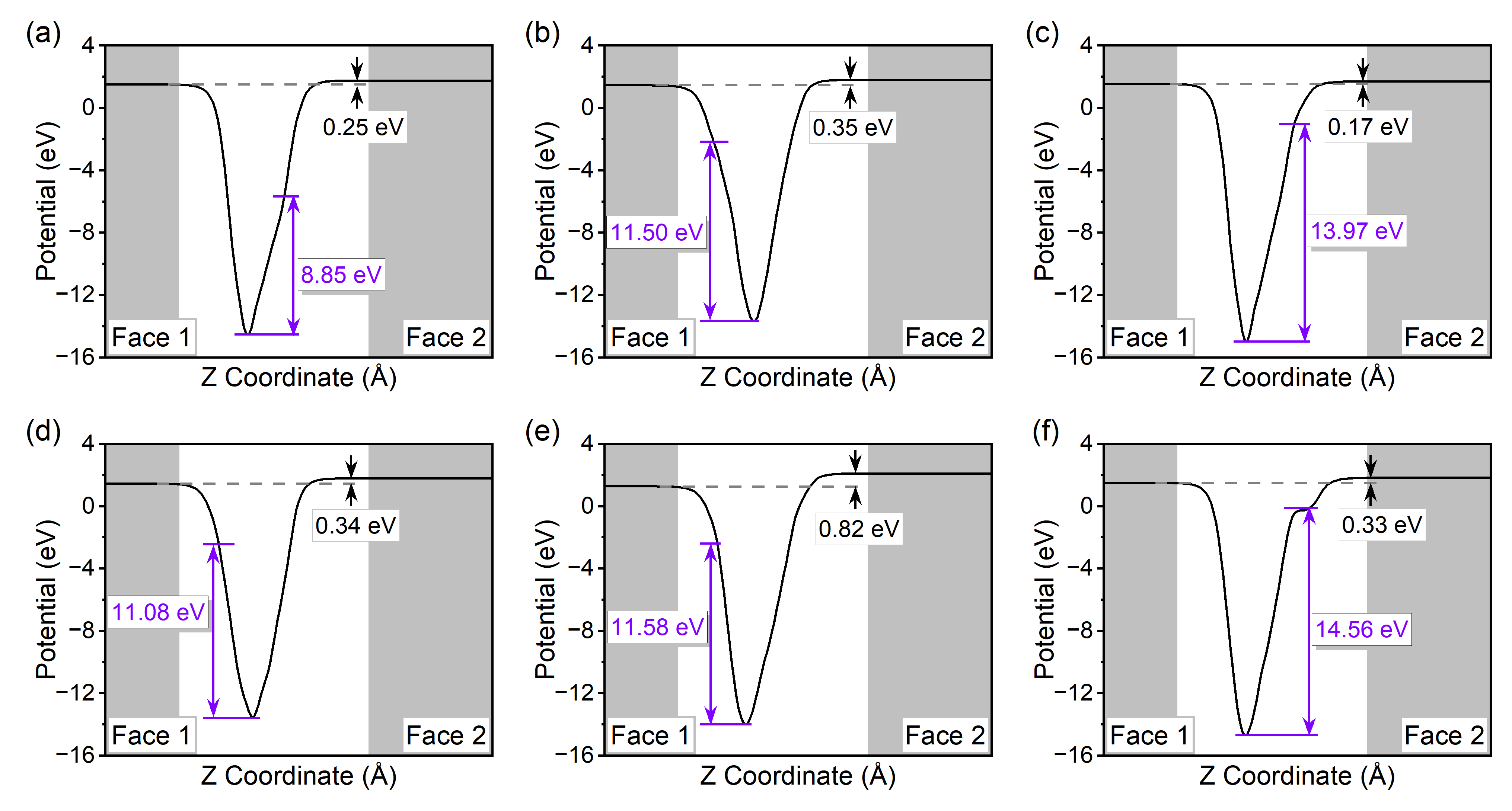}
	\caption {(a)-(f) The distribution of electrostatic potential for $P(N1)$, $P(N5)$, $P(N8)$, $P(C1)$, $P(I1)$, and $P(I2)$, respectively. We use the gray block to indicate the faces 1 and 2 of structures, the black arrows to denote the difference in vacuum levels on two surfaces, and the violet short lines and arrows to denote the electrostatic potential gap between the impurity site and the site of minimum value.} \label{fig5}
\end{center}
\end{figure}

Here we display the exact band structures of all selected doped structures in Fig.\ref{fig4}. In supporting information, we displayed all band structures as shown in Figure S4-S6. According to Fig.\ref{fig1}(e), both $N1$, $N5$, and $N8$ have high DOS occupation as similar, but $P(N8)$ has a shallow impurity level when $P(N1)$ and $P(N5)$ have deep cases (seeing in Fig.\ref{fig4}(a) to Fig.\ref{fig4}(c)). As mentioned above, we attribute it to the moving of impurity electrons induced by the electric field. To confirm the dependency relationship between the impurity level depth and the electrostatic potential change in P-C$_3$N$_5$, we analyzed the electrostatic potential distribution of $P(N1)$, $P(N5)$, and $P(N8)$ (seeing Fig.\ref{fig5}(a) to 5(c)).

As shown in Fig.\ref{fig5}(a) to Fig.\ref{fig5}(c), the difference in electrostatic potential between the dopant site and minimum value of P(N8) is huger than cases of $P(N1)$ and $P(N5)$. Due to that, impurity electrons in $P(N8)$ will move to the site of high potential, and the spatial density near the impurity atoms will decrease. Furthermore, the intrinsic electric field induced by the polarity has a similar direction to the electrostatic potential gradient between the dopant site and the minimum value. This will strengthen the movement of impurity electrons in $P(N8)$.

Because of high formation energies, $P(C2)$, $P(C3)$, $P(C4)$, and $P(C5)$ are not in the structures we screen out. However, the effects of the electrostatic field and DOS contribution of substituted atoms on the impurity level depth in $P(C1)$, $P(C2)$, $P(C3)$, $P(C4)$, and $P(C5)$ are still in approximate agreement with our predictions. Here we list the DOS integration of substituted carbon atoms in g-C$_3$N$_5$, the potential difference between the dopant site and the minimum value, the difference in electrostatic potential on two surfaces, and the depth of the impurity level for donor-doped P-C$_3$N$_5$ in Tab.\ref{tab1}. For donor doping structures, the depth of the impurity level is defined as the difference between its position and valence band maximum(VBM). 

\begin{table}[H]
\begin{center}
\caption{\label{tab1}%
The DOS integration of substituted carbon atoms in g-C$_3$N$_5$, the potential difference between the dopant site and the minimum value, the difference in electrostatic potential on two surfaces, and the depth of the impurity level for donor-doped P-C$_3$N$_5$. The negative value of the surface potential difference indicates that the direction of the polarization electric field is opposite to that of the electrostatic potential gradient.
}
\begin{ruledtabular}
\begin{tabular}{ccccc}
\textrm{~} &
\textrm{\makecell[c]{DOS\\Integration}} &
\textrm{\makecell[c]{Potential\\Difference(V)}} &
\textrm{\makecell[c]{Surface Potential\\Difference(eV)}} &
\textrm{\makecell[c]{Depth of\\Impurity Level(eV)}}\\
\colrule
$P(C1)$ & 0.65 & 11.08 & $-0.34$ & 1.16 \\
$P(C2)$ & 0.68 & 12.93 & $-0.57$ & 1.03 \\
$P(C3)$ & 0.57 & 14.01 & $-0.61$ & 1.81 \\
$P(C4)$ & 0.55 & 13.00 & 0.13 & 0.78 \\
$P(C5)$ & 0.57 & 4.32 & 0.10 & 1.43 \\
\end{tabular}
\end{ruledtabular}
\end{center}
\end{table}

As shown in Tab.\ref{tab1}, for $P(C3)$ and $P(C5)$, the DOS contributions of substituted carbon atoms are similar, and the greater potential difference between the dopant site and the minimum value in $P(C3)$ should lead the depth of the impurity level to be shallower than what is in $P(C5)$. However, the direction of the polarisation electric field in $P(C3)$ is opposed to the gradient of the potential difference so the polarity of $P(C3)$ will hinder the electron moving caused by the potential difference. Phosphorous doping almost doubles the difference between the vacuum levels of two surfaces in $P(C3)$. The magnified polarity leads to the reflux of impurity electrons and deepens the impurity level. Here we should note that the DOS occupation of the substituted atoms remains dominant in the depth of impurities, and the effect of the electrostatic potential gradient can only be apparent in the case of an approximation occupied by a substituted atomic DOS.

Generally, impurity levels induced into the band gap somehow weak the redox ability of photo-generated carriers\cite{pows4}. In this study, phosphorus doping magnifies the electrostatic potential difference between two surfaces of P-C$_3$N$_5$ and counteracts this weakness. Compared with the intrinsic structure, the electrostatic potential difference is enhanced in doping systems of $P(N5)$, $P(C1)$, $P(I1)$, and $P(I2)$(seeing Fig.\ref{fig5}), which indicates the improvement in the redox ability of photo-generated carriers in the catalyst. An obvious example is $P(I1)$: the regulated potential difference allows the material to carry out the HER in face 2, which is impossible in face 1 because of the poor reductive ability of photo-generated electrons. Furthermore, the magnification of the gap between VBM and the potential requirement of OER indicates the enhancement in the driving force of photo-generated holes used for OER. In the point of the band structure, $P(N1)$, $P(N5)$, $P(N8)$, and $P(I1)$ are suitable to be photocatalysts of POWS. The reduction capacity of photo-generated electrons in the $P(C1)$ structure cannot meet the requirements of HER, and the oxidation capacity of photo-generated holes in the $P(I2)$ is only a little higher than that required by OER, which means that $P(C1)$ and $P(I2)$ are only suitable to be the catalysts of the half-reactions in POWS but not the overall water splitting. In the next subsection, we will analyze the improvement in solar utilization brought about by the appearance of impurity levels.

\subsection{Solar absorption performance}

\begin{figure}[H]
\begin{center}
	\includegraphics[width=8cm]{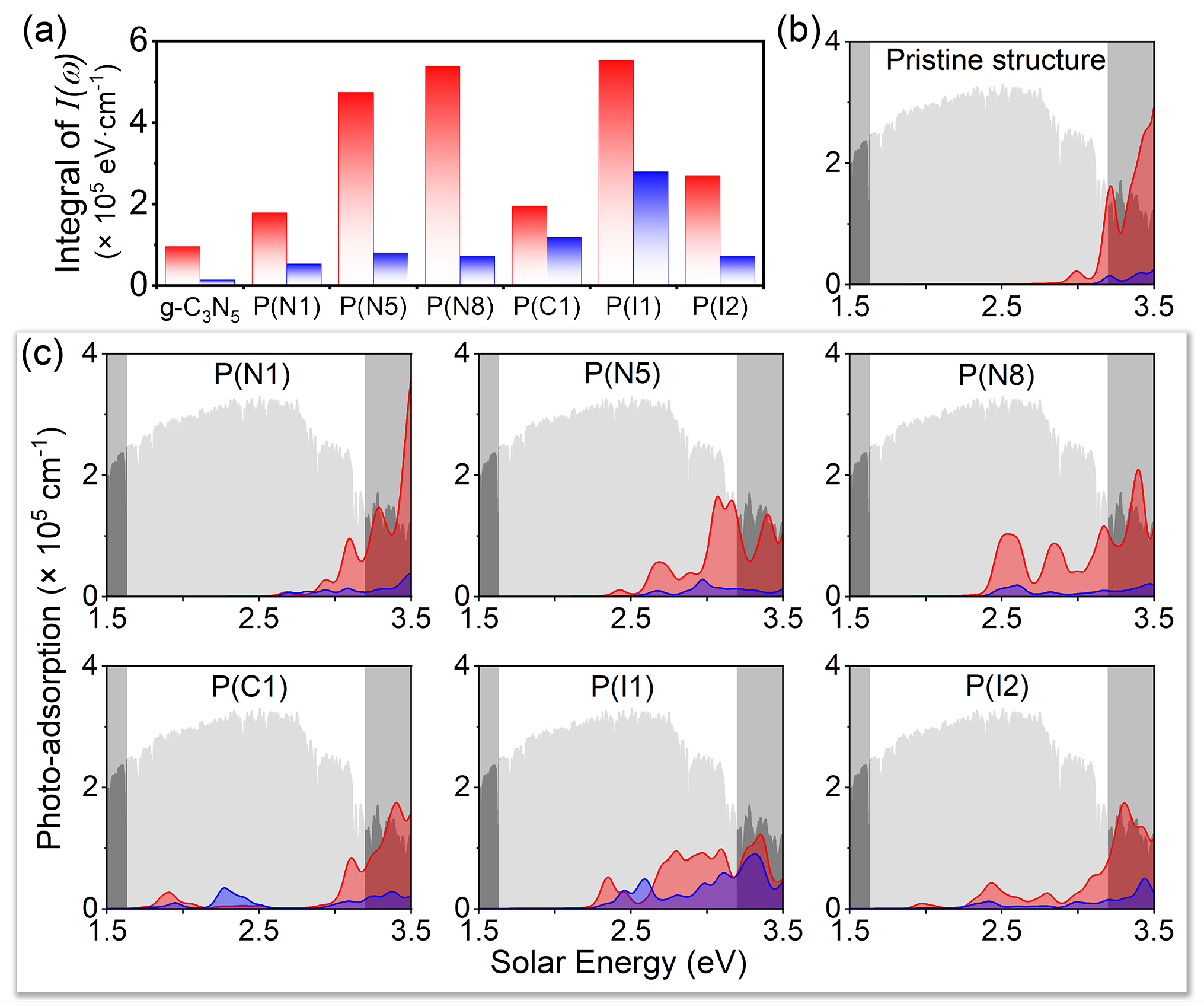}
	\caption {(a) The ability of the solar absorption for all selected structures. (b) The photo-adsorption coefficient of pristine g-C$_3$N$_5$. (c) The photo-adsorption coefficient of $P(N1)$, $P(N5)$, $P(N8)$, $P(C1)$, $P(I1)$, and $P(I2)$. We use the light gray background to show the intensity of the solar at different energy positions and clip out the energy range of the visible light with gray blocks, and the red and blue colors denote the absorption ability of materials with $x$ and $y$ directions light.} \label{fig6}
\end{center}
\end{figure}

A wide range of solar absorption is essential for the photocatalyst. In Fig.\ref{fig6}, we display the performance of the solar utilization for the pristine g-C$_3$N$_5$ and all selected doping types. Some structures have solar absorption with high intensity but a narrow absorption range, so it brings an imperfect result to discuss the ability of a material to utilize sunlight purely in terms of the height of the absorption peak or the width of the absorption spectrum. Herein, we calculate the integration of the photo-absorption coefficient in the visible region to characterize the ability of the solar absorption for all structures. Because the solar intensity is higher in the interval of visible light, in which the region is from 1.64 eV to 3.19 eV, our integral range only covers this area.

In our results, all doping structures expand their region of photo-absorption spectrum because of the induction of impurity levels. In the visible light region, the absorption peaks of all materials are more intense for $x$ direction polarized light than for $y$ direction polarized light, corresponding to a broader absorption spectrum width also(seeing Fig.\ref{fig6}(c)). This may be related to the inclination angle of each cell for g-C$_3$N$_5$ systems: the inclination direction of the normal in the cell plane is the same as the polarization direction of the light with higher absorbance. The most obvious expansion of the optical absorption spectrum is in the $P(C1)$ and $P(I2)$ doped structures, where the absorption peaks appear in the range of photon energies less than 2 eV. However, the absorption spectrum ranges of $P(C1)$ and $P(I2)$ are wide but with weak intensity in the region of intenser solar illumination. The $P(I1)$ doping structure has the best performance of photo-absorption in the region of higher solar intensity for either $x$ or $y$ direction polarization. Combined with the previous analysis on the edge level of band structures, it appears that $P(I1)$ will be the best structure to be the catalyst of the POWS among the selected doping models.

\subsection{Simulation of POWS}

\begin{figure}[H]
\begin{center}
	\includegraphics[width=8cm]{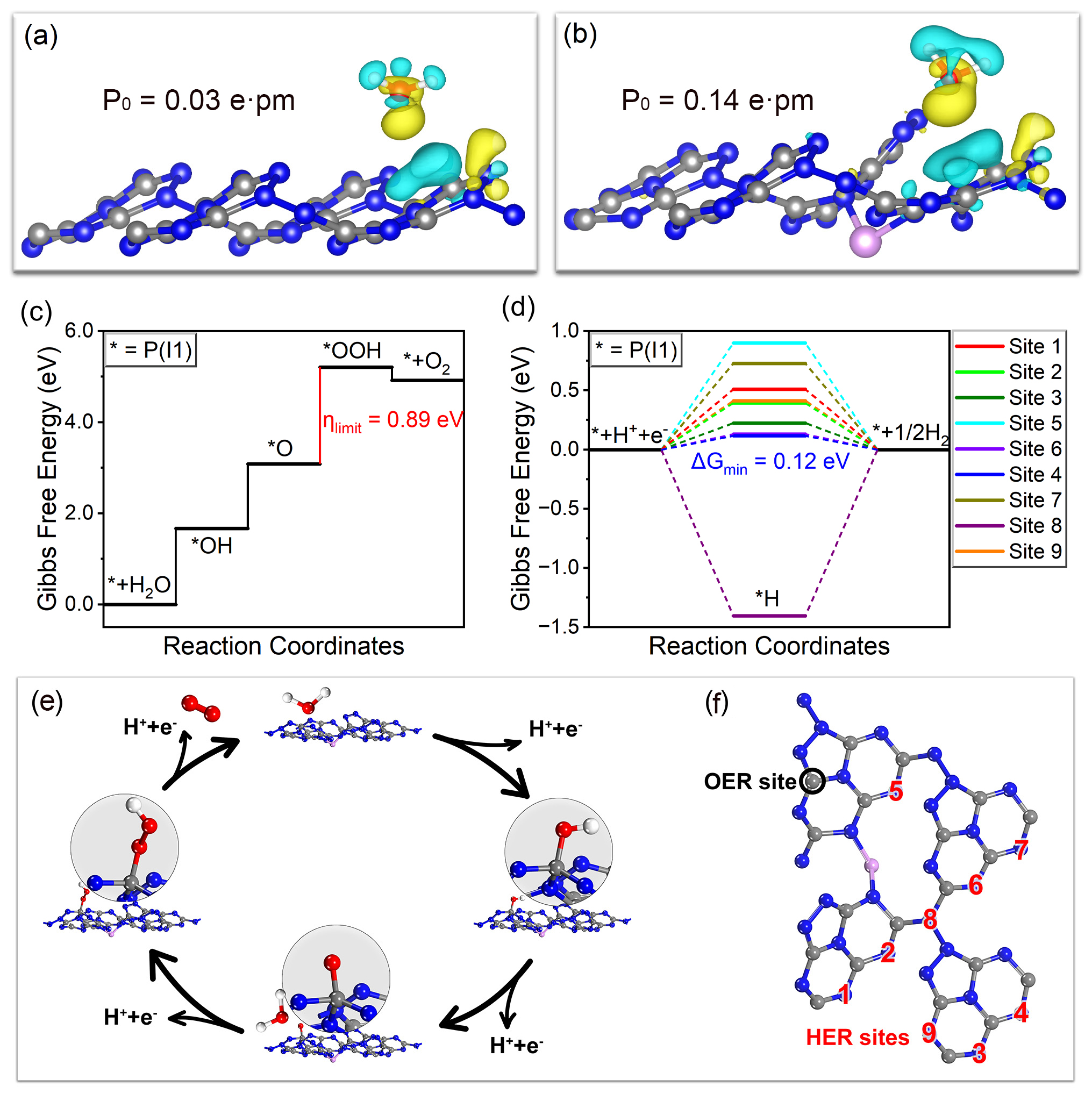}
	\caption {(a) $\&$ (b) The change difference before and after the adsorption of the water molecule on the catalyst and the corresponding overall surface dipole moment of pristine g-C$_3$N$_5$ and $P(I1)$, respectively. (a) is for pristine g-C$_3$N$_5$, and (b) is for $P(I1)$. The value of isosurfaces is set to 0.0005 e/Bohr$^3$. (c) $\&$ (d) The variations of Gibbs free energy in the process of OER and HER, respectively. We use the red color to denote the energy barrier and the over-potential of the limit step. (e) The scheme of OER process. We zoomed in on the structural details of the adsorption section. (f) The sites of OER and HER. We use the black circle to highlight the OER site, and red numbers are used to label different sites of the HER simulation. The gray, blue, light purple, red, and white balls represent carbon, nitrogen, phosphorus, oxygen, and hydrogen atoms, respectively. The yellow and cyan areas represent the regions of electron dissipation and accumulation.} \label{fig7}
\end{center}
\end{figure}

The process of POWS can be divided into three parts: the adsorption of the water molecule, the OER process, and the HER process. Here we first compare the adsorption of the water molecule in $P(I1)$ with the case in the pure structure. The electric field of the intrinsic surface dipole can enhance the charge transfer between the target molecular and catalyst, then the absorption and subsequent splitting of water molecules can be promoted. Since the beginning of the POWS is the out-of-plane adsorption of the water molecular, we calculated the dipole distribution of the catalyst in the out-of-plane direction, i.e., the  direction. The surface dipoles for pristine g-C$_3$N$_5$ and $P(I1)$ are displayed in Fig.\ref{fig7}(a) and Fig.\ref{fig7}(b). Compared to cases in pure g-C$_3$N$_5$, the doping strategy enhances the overall dipole of $P(I1)$. A higher dipole movement brings a stronger interaction between the catalyst and target molecules, which can be reflected by the charge difference distribution in Fig.\ref{fig7}(b). Due to that, there is a lower absorption energy ($-$0.25 eV) relative to the case of the pure system ($-$0.18 eV) when the water molecule is adsorbed to the catalyst.

To analyze the reliability and the activity of the $P(I1)$ to be a catalyst of POWS, we performed the simulation for the process of OER and HER with an acidic condition. The OER process is set as the classical four electronic steps, and the limit step is identified as the one with the highest barrier of Gibbs free energy. As shown in Fig.\ref{fig7}(c), the limit step of OER is the third step, and the corresponding over-potential of 0.89 eV can be stridden over thermodynamically by the strong driving force of photo-generated holes because of the wide gap between the VBM and the requirement potential of OER in $P(I1)$ structures. Furthermore, we simulated HER on $P(I1)$. The lowest free energy barrier of 0.12 eV proves that photocatalytic HER based on the $P(I1)$ structure is also feasible.

\section{Conclusions}
In this paper, we investigated the application of KL-EMT in the doping strategy by doping the phosphorus atom in the g-C$_3$N$_5$. We find that the positions of impurity levels not only depend on the projected density of states (pDOS) of substituted atoms in the pure g-C$_3$N$_5$ but also rely on the electrostatic potential gradient and polarity induced by phosphorus doping. Moreover, impurity levels in the band gap expand the solar absorption range of g-C$_3$N$_5$, enhanced surface dipole and redox ability of photo-generated carriers significantly promote adsorption and subsequent splitting of water molecules. POWS simulating and analysis of Gibbs free energy confirm the potential of P-C$_3$N$_5$ to be the catalyst of POWS. Our study extends the application context of KL-EMT and proposes a viable strategy for the design of POWS catalysts.

\begin{acknowledgments}
This work is supported by China Postdoctoral Science Foundation (Grant No. 2022M711691), National Natural Science Foundation of China (Grant No. 12104130, 12047517), Six talent peaks project in Jiangsu Province (Grant No. XCL-104), Postgraduate Research $\&$ Practice Innovation Program of Jiangsu Province (Grant No. KYCX22$\_$0901, KYCX22$\_$0991).
\end{acknowledgments}

\appendix

\section{Simulation of Water Splitting}
For the process of overall water splitting, we consider it as two half-reactions: the hydrogen evolution reaction (HER) and the oxygen evolution reaction (OER). The reaction environment is assumed to be acidic.
\subsection{Hydroygen Evolution Reaction (HER)}
The process of HER is the classical single electron step process\cite{ling}:
\begin{equation}
    * + \left(H^{+}+e^{-}\right) \rightarrow *H 
\end{equation}
where $*$ represents the catalyst. The corresponding Gibbs free energy change is
\begin{equation}
    \Delta G = G\left(*H\right) - G\left(*\right) -G\left(H^{+}+e^{-}\right) 
\end{equation}
where $G(X)$ represents the Gibbs free energy of $X$. $G\left(H^{+}+e^{-}\right)$ is calculated by the computational hydrogen electrode (CHE) model\cite{che}:
\begin{equation}
    G\left(H^{+}+e^{-}\right) = \dfrac{1}{2} G\left(H_{2}\left(gas\right)\right)
\end{equation}
where $H_{2}\left(gas\right)$ represents the hydrogen molecule in the gas phase.

\subsection{Oxygen Evolution Reaction (OER)}
The process of OER is the classical four electron step process. The processes and corresponding Gibbs free energy change are\cite{oer}:\\
\noindent First:
\begin{equation}
    * + H_{2}O \rightarrow *OH + \left(H^{+}+e^{-}\right)
\end{equation}
\begin{equation}
    \Delta G_{1} = G\left(*OH\right) + G\left(H^{+}+e^{-}\right) - G\left(*\right) -G\left(H_{2}O\right)
\end{equation}
Second:
\begin{equation}
    *OH \rightarrow *O + \left(H^{+}+e^{-}\right)
\end{equation}
\begin{equation}
    \Delta G_{2} = G\left(*O\right) + G\left(H^{+}+e^{-}\right) - G\left(*OH\right) 
\end{equation}
Third:
\begin{equation}
    *O + H_{2}O \rightarrow *OOH + \left(H^{+}+e^{-}\right) 
\end{equation}
\begin{equation}
    \Delta G_{3} = G\left(*OOH\right) + G\left(H^{+}+e^{-}\right) - G\left(*O\right) -G\left(H_{2}O\right)
\end{equation}
Fourth:
\begin{equation}
    *OOH \rightarrow * + O_{2}\left(gas\right) + \left(H^{+}+e^{-}\right) 
\end{equation}
\begin{equation}
    \Delta G_{4} = G\left(*\right) + G\left(H^{+}+e^{-}\right) + G\left(O_{2}\left(gas\right)\right) - G\left(*\right)
\end{equation}
where $O_{2}\left(gas\right)$ represents the oxygen molecule in the gas phase. The limiting step is chosen to be the step with the largest of the Gibbs free energy changes:
\begin{align}
    \Delta G_{max} = \left\{ Max \left| \Delta G_{1}, \Delta G_{2}, \Delta G_{3}, \Delta G_{4} \right. \right\} 
\end{align}
The overpotential $\eta$ of OER is calculated by
\begin{align}
    \eta = \dfrac{\Delta G_{max}-1.23 eV}{e}
\end{align}
where $e$ is the absolute value of the electron charge.

\section{Calculation of Gibbs Free Energy}
The Gibbs free energy is defined as\cite{gibbs}
\begin{align}
    G=E+ZPE-TS 
\end{align}
where $E$ is the energy based on DFT calculations, $ZPE$ is the zero point energy correction, $T$ is the temperature, and $S$ is the entropy. $ZPE$ is defined as 
\begin{equation}
    ZPE=\dfrac{1}{2}\sum_{i}h\nu 
\end{equation}
$S$ is defined as 
\begin{align}
    S = \sum_{i} R &\left\{ \dfrac{h\nu_{i}}{k_{B}T} \left[ \dfrac{h\nu_{i}}{k_{B}T} - 1 \right]^{-1}  - \ln \left[ 1 - \exp \left( -\dfrac{h\nu_{i}}{k_{B}T} \right) \right] \right\}    
\end{align}
where $h$, $R$, $k_{B}$ and $\nu_{i}$ are Planck constant, Gas constant, Boltzmann constant and the $i$-th vibration frequency. $i$ is the frequency number generally taken from 1 to 3$N$, where $N$ is the number of atoms vibrating.

\nocite{*}
\bibliography{aipsamp}

\end{document}